\documentclass[twocolumn,showpacs,preprintnumbers,amsmath,amssymb,superscriptaddress]{revtex4}

\usepackage{graphicx}
\usepackage{epstopdf}
\usepackage{dcolumn}
\usepackage{bm}

\begin{document}


\title{ 
Hg-Based Superconducting Cuprates: 
High $T_\mathrm{c}$ and Pseudo Spin-Gap 
}

\author{Y. Itoh}

\address{Department of Chemistry, Graduate School of Science, 
Kyoto University, Kyoto 606-8502, Japan}

\author{T. Machi}

\address{Superconductivity Research Laboratory,
 International Superconductivity Technology Center,
1-10-13 Shinonome, Koto-ku, Tokyo 135-0062, Japan}

\date{\today}%

\begin{abstract}
Mercury-based cuprates HgBa$_2$CuO$_{4+\delta}$ with 0 $< \delta <$ 0.2 (Hg1201) are the superconductors with a single CuO$_2$ layer in unit cell and the optimally oxidized one has the highest $T_\mathrm{c}$ = 98 K among the ever reported single-CuO$_2$-layer superconductors. 
Double CuO$_2$ layered cuprates HgBa$_2$CaCu$_2$O$_{6+\delta}$ with 0.05 $< \delta <$ 0.35 (Hg1212) have the highest $T_\mathrm{c}$ = 127 K at the optimal oxygen concentration.
This is  the highest $T_\mathrm{c}$ among the ever reported double-CuO$_2$-layer superconductors.  
The Hg1201 has the nearly perfect fat CuO$_2$ plane. 
The Hg1212 has the flattest CuO$_2$ plane among the other lower $T_\mathrm{c}$ double-layer cuprates, which is associated with the mystery of the highest $T_\mathrm{c}$. 
Both systems have a pseudo spin-gap in the magnetic excitation spectrum of the normal states. 
In this article, we present the microscopic studies of magnetic and electric properties of the Hg-based superconducting cuprates using nuclear magnetic resonance (NMR) and nuclear quadrupole resonance (NQR) techniques. NMR and NQR are powerful to detect local information through the nuclear sites in materials and have supplied us with information on low frequency magnetic response of electronic systems. 
Although the structure analysis indicates the flat CuO$_2$ planes, zero field $^{63, 65}$Cu NQR spectra, which are sensitive to local electric charge distribution, show inhomogeneous broadening. The local electrostatic states are rather inhomogeneous. 
Although the $d$-wave superconductivity must be fragile to imperfection and non-magnetic impurities, 
the pure Hg-based superconducting cuprates show impure $^{63, 65}$Cu NQR spectra 
but rather robust pseudo spin-gap in the $^{63}$Cu NMR Knight shift and nuclear spin-lattice relaxation rate over the wide doping regions. 
There had been an issue whether the pseudo spin-gap results from a double-layer coupling
or a single layer anomaly.
The NMR results for Hg1201 served as the evidence for the existence of the single-layer pseudo spin-gap.  
The pseudo spin-gap is explained by a precursory phenomena of superconducting pairing fluctuations or spin singlet correlation. The different doping dependence of the pseudo spin-gap of Hg1201 and Hg1212 is associated with the different Fermi surface contour.
The similar temperature dependence of the $^{199}$Hg and the $^{63}$Cu nuclear spin-lattice relaxation times  indicates uniform interlayer coupling. 
The double-layer coupling effect is revisited through the comparison of Hg1201 and Hg1212.  
 
\end{abstract}

\maketitle

\tableofcontents

\section{INTRODUCTIN}
\label{sec:intro} 
The BCS theory is one of the most successful theories in physics.
It has not only revealed the mechanism of phonoic superconductivity
but also developed our understanding of spontaneous breaking of gauge symmetry. 
The BCS theory as a field theory has been applied to various stages of physics, elementary particles, condensed matter and cosmology. 

The discovery of high-$T_\mathrm{c}$ cuprate superconductors has renewed our interests of superconductivity.
Low dimensionality on layered compounds, carrier doping effect, electron correlation effect, antiferromagnetic ordering and Mott transition 
play the key roles to understand the high-$T_\mathrm{c}$ physics. 
Since the high-$T_\mathrm{c}$ superconductivity emerges in close to antiferromagnetic instability,
the significant role of magnetism is associated with the high-$T_\mathrm{c}$ mechanism. 
Since Coulomb repulsion between electrons can produce spin and charge fluctuations,
a close relation between magnetism and superconductivity is suggested. 
The superconducting properties except the high $T_\mathrm{c}$ and vortex matter physics are conventional. 
The electronic and magnetic properties in the normal conducting state are unconventional.  
The high-$T_\mathrm{c}$ cuprate superconductors have still attracted great interests.   
We may anticipate new development in understanding the solid state physics.  
  
Mercury-based cuprates HgBa$_2$CuO$_{4+\delta}$ with 0 $< \delta <$ 0.2 (Hg1201) are the superconductors with a single CuO$_2$ layer in unit cell and the optimally oxidized one has the highest $T_\mathrm{c}$ = 98 K among the ever reported single-CuO$_2$-layer superconductors~\cite{Antipov1,Fukuoka1}. 
Mercury-based cuprates HgBa$_2$CaCu$_2$O$_{6+\delta}$ with 0.05 $< \delta <$ 0.35 (Hg1212) are the superconductors with double CuO$_2$ layers in unit cell and the optimally oxidized one has the highest $T_\mathrm{c}$ = 127 K among the ever reported double-CuO$_2$-layer superconductors~\cite{Ott}. 
Hg1201 and Hg1212 are the layered compounds and contain one and two CuO$_2$ planes in unit cell, respectively~\cite{Antipov2}.  
Triple-CuO$_2$-layer HgBa$_2$Ca$_2$Cu$_3$O$_{8+\delta}$ (Hg1223) is the highest $T_\mathrm{c}$ = 134 K at ambient pressure~\cite{Ott} and $T_\mathrm{c}$ = 164 K at high pressure~\cite{Chu}.
Hg1234 is also known to be synthesized~\cite{AntipovHg1234}.
 But it has lower $T_\mathrm{c}$ than Hg1223~\cite{Usami}. 
 
In this article, we present the Hg-based superconducting cuprates and  
the microscopic magnetic properties studied through nuclear magnetic resonance (NMR) and nuclear quadrupole resonance (NQR) techniques. 
Especially, we present the Cu NMR evidence of the existence of a pseudo spin-gap in the norma-state magnetic excitation spectrum
of the single-layer cuprate superconductor. 
The discovery of a big pseudo spin-gap of the underdoped single-layer cuprate superconductor
has turn our attention to the unconventional electronic state above $T_\mathrm{c}$.     

\section{PSEUDO SPIN-GAP: SECONDARY OR INHERENT?}
Conventional metallic states are well described by the Landau-Fermi liquid theory.  
Uniform magnetic susceptibility $\chi_s$ exhibits Pauli paramagnetism.  
Since the density of electron states is finite at the Fermi level, 
$\chi_s$ is finite at $T$ = 0 K and nearly independent of temperature. 
Nuclear spin-lattice relaxation time $T_1$ satisfies Korringa relation with the spin Knight shift.   
Since the quasi-particle scattering  through the finite density of states induces the nuclear spin relaxation,
1/$T_{1}T$ is finite at $T$ = 0 K.  

Magnetic itinerant compounds involving the transition metal elements 
often exhibit Curie-Weiss magnetism, in spite of the absence of localized moments. 
Itinerant magnetism breaks down the Korringa relation. 
In antiferromagnetic compounds, the nuclear spin-lattice relaxation rate  
divided by temperature 1/$T_1T$
is enhanced in a Curie-Weiss law, more than the uniform spin susceptibility $\chi_s$. 
 
The normal-state pseudo spin-gap was first confirmed by NMR experiments~\cite{ImaiPSG,YasuokaPSG,AOM,Warren,Horvatic,Takigawa}.  
The decrease of a static uniform spin susceptibility $\chi_\mathrm{s}$ with cooling down was observed in the NMR Knight shift measurements
for the widely-studied double-layer YBa$_2$Cu$_3$O$_{7-\delta}$. 
The decrease of the nuclear spin-lattice relaxation rates divided by temperature 1/$T_1T$
was also observed for the underdoped cuprates in the normal state. 
Thus, the existence of a pseudogap in the  magnetic excitation spectrum
was observed in the underdoped double-layer superconductors in the normal states. 
However, the existence of such a pseudogap  in the  magnetic excitation spectrum was not obvious for the single-layer superconductor La$_{2-x}$Sr$_x$CuO$_{4}$. Then, there had been an issue whether the pseudogap  in the  magnetic excitation spectrum is intrinsic in a CuO$_2$ plane
or it is secondary due to spin singlet formation between adjacent CuO$_2$ planes in the double layer.      
The NMR studies of Hg1201 gave us an obvious evidence of the pseudo spin-gap in the single layer. 

The review articles of the intensive NMR studies of the high-$T_\mathrm{c}$ YBa$_2$Cu$_3$O$_{7-\delta}$ can be seen in~\cite{PS,Berthier,Russ,ItohP}. 
Hence, we focus on the NMR studies of Hg1201 and Hg1212. 

\section{CRYSTAL STRUCTURE AND PHASE DIAGRAM}
\subsection{Record high $T_\mathrm{c}$}
\begin{figure}
\begin{center}
\includegraphics[width=0.8\linewidth]{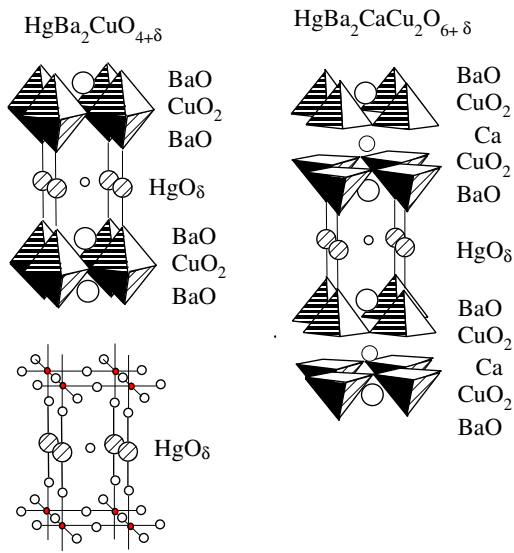}
\end{center}
\caption{
Crystal structures of single-layer HgBa$_2$CuO$_{4+\delta}$
and double-layer HgBa$_2$CaCu$_2$O$_{6+\delta}$. 
Strong O-Hg-O bonds like ``dumbbell" along the $c$-axis and dilute oxygen concentration
in the HgO layers are characteristics of these crystal structures. 
}
\label{CS}
\end{figure} 
\begin{figure}
\begin{center}
\includegraphics[width=0.7\linewidth]{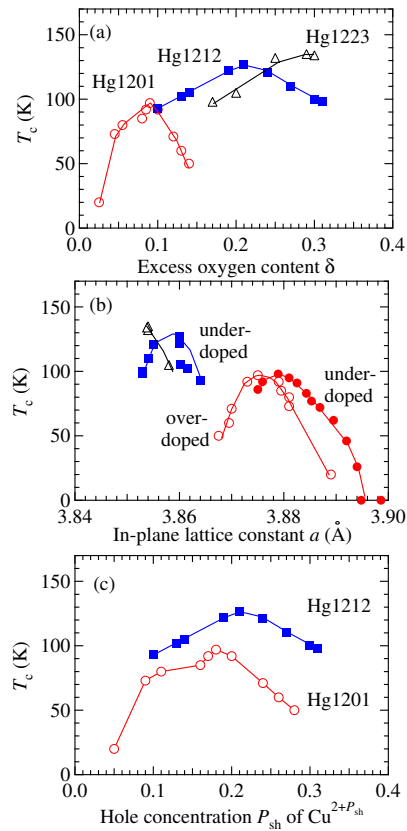}
\end{center}
\caption{
$T_\mathrm{c}$ phase diagrams of Hg1201 and Hg1212 reproduced from~\cite{Fukuoka2,YamamotoHg1201}. 
$T_\mathrm{c}$ is plotted against excess oxygen concentration $\delta$ ~\cite{Fukuoka2}(a), the in-plane $a$-axis lattice constant ~\cite{Fukuoka2,YamamotoHg1201}(b), and hole concentration $P_\mathrm{sh}$ defined by the ionic formal valence of Cu$^{2+P_\mathrm{sh}}$ ($P_\mathrm{sh}$ = 2$\delta$ for Hg1201 and $P_\mathrm{sh}$ = $\delta$ for Hg1212) ~\cite{Fukuoka2}(c).
}
\label{PD}
\end{figure}  
Figure~\ref{CS} shows the crystal structures of the Hg-based superconductors Hg1201 and Hg1212~\cite{Antipov2}. 
Strong covalent O-Hg-O bonds along the $c$-axis, which look like ``dumbbell" shape, and dilute oxygen concentration in the HgO$_{\delta}$ layers are characteristics of the crystal structures. 
The change in oxygen concentration of the HgO$_{\delta}$ layers can yield a wide carrier doping region.  

Figure~\ref{PD} (a) shows the oxygen concentration dependences of $T_\mathrm{c}$ of
Hg1201, Hg1212, and Hg1223~\cite{Fukuoka2}. 
The typical ``bell"-shaped dependences of $T_\mathrm{c}$ are seen for Hg1201 and Hg1212. 
With changing only the oxygen concentration, 
the electronic states of Hg1201 and Hg1212 develop from the underdoped to the overdoped regions
Although no magnetic ordering states nor insulating states were confirmed 
for the deeply underdoped samples of Hg1201 and Hg1212, 
the existence of Hg-based cuprate insulators was observed for Hg$_2$Ba$_2$YCu$_2$O$_{8-\delta}$ (Hg2212) 
which contains the double Hg layers~\cite{Radaelli1,Radaelli2,YamamotoHg2212}. 

Figure~\ref{PD} (b) shows $T_\mathrm{c}$ plotted against the in-plane $a$-axis lattice parameter for
Hg1201, Hg1212, and Hg1223~\cite{Fukuoka2,YamamotoHg1201}. 
The in-plane Cu-Cu distances of Hg1212 and Hg1223 
are shorter than that of Hg1201. 
The shrunk CuO$_2$ planes characterize the higher $T_\mathrm{c}$ cuprates.  

Figure~\ref{PD} (c) shows $T_\mathrm{c}$ plotted against the hole concentration
$P_\mathrm{sh}$ defined by the ionic formal valence of Cu$^{2+P_\mathrm{sh}}$~\cite{Fukuoka2}.
We estimated $P_\mathrm{sh}$ = 2$\delta$ for Hg1201 and $P_\mathrm{sh}$ = $\delta$ for Hg1212. 
The  optimal hole concentration is nearly the same as the typical values of 0.18-0.20.
Then, the mechanism of the record high $T_\mathrm{c}$ is not only due to the substantial hole concentration.  

 \subsection{Flat CuO$_2$ plane}
\begin{figure}
\begin{center}
\includegraphics[width=0.9\linewidth]{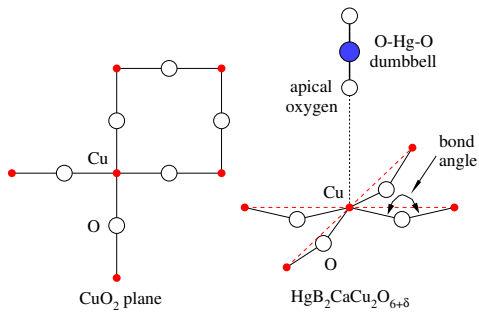}
\end{center}
\caption{
A CuO$_2$ plane (left) and two key structure parameters (right) are illustrated.
The two key structure parameters (right) are the distance from the plane-site Cu
to the apical oxygen and 
the bond angle between the Cu ions via the oxygen ion in the CuO$_2$ plane. 
The flatness of the CuO$_2$ plane is believed to be a significant factor 
to the higher $T_\mathrm{c}$.  
}
\label{CuO2}
\end{figure}

Figure~\ref{CuO2} illustrates a CuO$_2$ plane and two key structure parameters. 
The two key parameters are the distance from the plane-site Cu ion 
to the apical oxygen and the bond angle between the plane-site Cu ions via the plane-site oxygen. 
Both parameters are associated with the flatness of the CuO$_2$ planes. 
The large separation between Cu and the apical oxygen and the nearly 180 degree of the in-plane Cu-O-Cu bond angle are observed in Hg1201 and Hg1212~\cite{Wanger1,Antipov3,Antipov4,Huang,Wagner2,Radaelli3}. 
The optimal $T_\mathrm{c}$ = 38 K La$_{2-x}$Sr$_x$CuO$_4$ with the single CuO$_2$ layers and
the optimal $T_\mathrm{c}$ = 93 K YBa$_2$Cu$_3$O$_{6.9}$ with the double CuO$_2$ layers 
have the buckling structures in the CuO$_2$ planes. 
The CuO$_2$ planes of Hg1201 and Hg1212 are the flattest among the reported superconducting cuprates~\cite{Wanger1,Antipov3,Antipov4,Huang,Wagner2,Radaelli3}.

Many researchers believe that the nearly perfect flat CuO$_2$ plane is a significant factor to realize the higher $T_\mathrm{c}$. 
Actually, the double-Hg-layer (Hg, Tl)2212 has the relatively lower $T_\mathrm{c}$
and involves the bucking in the CuO$_2$ planes~\cite{YamamotoHg2212,Ohta}.  
In the (Hg, Tl)2212, the in-plane Cu-O-Cu bond is bended and the bond angle is about 170 degree.   
The flatness of the CuO$_2$ planes is associated with a higher  $T_\mathrm{c}$ mechanism or at least an inevitable background of crystal structure.

\section{TWO DIMENSIONAL CONDUCTOR}
\begin{figure}
\begin{center}
\includegraphics[width=0.8\linewidth]{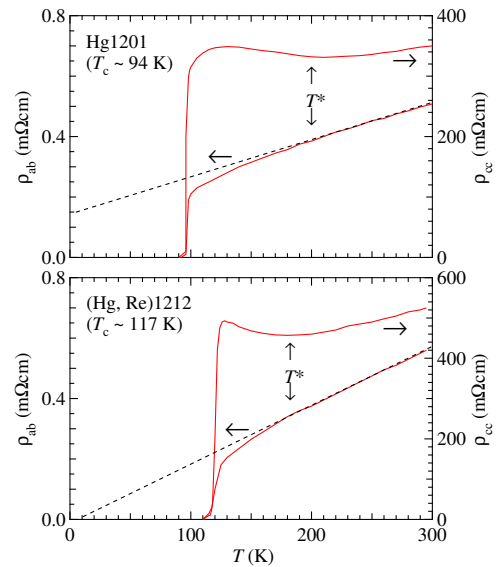}
\end{center}
\caption{
Anisotropic electrical resistivity of $\rho_{ab}$ and $\rho_{cc}$ of Hg1201 (upper panel)~\cite{Hardy} and Hg1212 (lower panel)~\cite{Ogawa1,Ogawa2}. 
Dash lines for the in-plane resistivity are $T$ linear functions. 
The onset temperatures of deviation from the $T$ linear functions are denoted by $T^*$,
which are also the minimum temperatures of the out-of-plane resistivity. 
}
\label{rhoT}
\end{figure}

Anisotropy of electrical resistivity has been measured for Hg1201 single crystals~\cite{Hardy} and 
Hg1212 epitaxial thin films on vicinal substrates~\cite{Ogawa1,Ogawa2}.
Figure~\ref{rhoT} shows the in-plane resistivity $\rho_{ab}$ and the out-of-plane resistivity $\rho_{cc}$ for Hg1201 single crystal of $T_\mathrm{c}$ = 97 K reproduced from~\cite{Hardy} (a)
and those of (Hg, Re)1212 of  $T_\mathrm{c}$ = 117 K reproduced from~\cite{Ogawa1,Ogawa2} (b).

Dash lines are $T$ linear functions for the in-plane resistivity $\rho_{ab}$. 
The $T$-linear resistivity is associated with the conduction electron scattering due to two dimensional antiferromagnetic spin fluctuations~\cite{MTU,YanaseR}.
The deviation from the $T$-linear behavior is associated with the scattering suppression due to the opening of a pseudo spin-gap in the magnetic excitation spectrum.
The onset temperatures of deviation from the $T$ linear functions are denoted by $T^*$.  

At $T^*$, the out-of-plane resistivity $\rho_{cc}$ also takes the minimum value. 
The out-of-plane resistivity $\rho_{cc}$ is two or three order higher than the in-plane resistivity $\rho_{ab}$ and the temperature dependence is different from each other.
The metallic in--plane resistivity $\rho_{cc}$ and the semi-conducting out-of-resisitivity $\rho_{cc}$
above $T^*$ is an evidence of two dimensional electrical conduction in Hg1201 and Hg1212.  
This has been known for the other superconducting cuprates.

The in-plane and the out-of-plane electrical conduction is correlated with each other
with respect to the $T^*$.   
The two dimensional electrical resistivity is understood in terms of 
a hot spot and a cold spot on the two dimensional Fermi surface~\cite{YanaseR}. 

\section{Cu NQR, NMR AND MAGNETIC FLUCTUATIONS}
Nuclear magnetic resonance (NMR) and nuclear quadrupole resonance (NQR) 
are powerful techniques to characterize microscopically magnetic insulators, metals, superconductors, alloys and compounds~\cite{Slichter}.
Microscopic studies using the NMR and NQR techniques have provided us with rich information  
inside unit cell through the nuclear sites in a site-selective way.  
Using the NMR and NQR techniques, one can obtain static and dynamic information of the electronic systems.   

\subsection{Inhomogeneous Cu NQR spectrum of pure Hg-based cuprate}
\begin{figure}
\begin{center}
\includegraphics[width=1.0\linewidth]{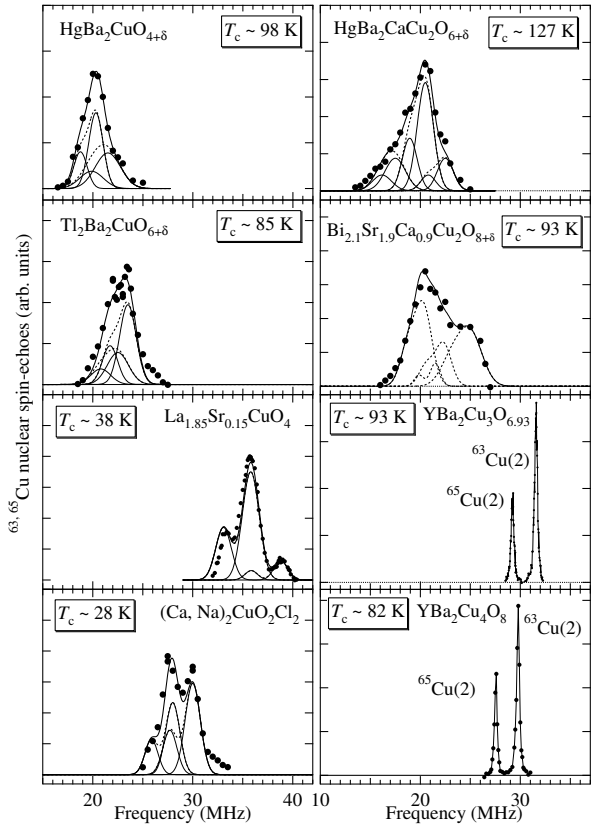}
\end{center}
\caption{
Zero-field plane-site $^{63, 65}$Cu NQR frequency spectra for single-CuO$_2$-layer  superconducting cuprates (left panels) (a) and double-CuO2-layer superconducting cuprates (right panels) (b) at $T$ = 4.2 K. 
}
\label{CuNQR}
\end{figure}  

$^{63, 65}$Cu nuclei have spin $I$ = 3/2 and then quadrupole moments $Q_m$. 
In a non-cubic crystalline symmetry, they interact with the electric field gradient $V_\mathrm{\alpha\beta}$ ($\alpha, \beta$ = $x, y, z$) of electric static crystalline potential $V$~\cite{Slichter}.  
Nuclear quadrupole Hamiltonian is given by
\begin{equation}
H_\mathrm{Q}={e^2qQ_m \over 4I(2I-1)}[(3I_z^2-I^2)+{\eta \over 2}(I_{+}^2+I_{-}^2)] 
\label{NQ}
\end{equation}
where $q$ is the maximum component of the electric field gradient
\begin{equation}
eq \equiv {\partial^2 V\over \partial z^2}
\label{Vzz}
\end{equation}
and $\eta$ is called the asymmetry parameter 
\begin{equation}
\eta \equiv {V_{xx}-V_{yy} \over V_{zz}}.
\label{eta}
\end{equation}
$^{63, 65}$Cu nuclear quadrupole resonance frequency $\nu_\mathrm{NQR}$ is given by
\begin{equation}
\nu_\mathrm{NQR}=\nu_{Q}\sqrt{1+{\eta^2 \over 3}},
\label{eta}
\end{equation}
where $\nu_{Q}$ = $e^2qQ_m$/2. 
The ratio of the natural abundance of $^{63}$Cu and $^{65}$Cu atoms is about 69.1 to 30.9.
The ratio of the nuclear quadrupole moments $Q_m$ is about 0.211 to 0.195.
For one crysallographic Cu site, 
a pair of $^{63}$Cu and $^{65}$Cu NQR lines is observed. 

Figure~\ref{CuNQR} shows actual zero-field plane-site $^{63, 65}$Cu NQR frequency spectra 
for various optimally doped single-CuO$_2$-layer  superconducting cuprates (left panels) (a) and double-CuO2-layer superconducting cuprates (right panels) (b) at $T$ = 4.2 K. 
The Cu NQR spectra were measured for Hg1201 in~\cite{ItohHg1201L,ItohHg1201F}, 
Tl$_2$Ba$_2$CuO$_{6+\delta}$~\cite{ItohTl2201P}, La$_{1.85}$Sr$_{0.15}$CuO$_4$ in~\cite{Yamagata}, (Ca, Na)$_2$CuO$_2$Cl$_2$ in~\cite{ItohCNCO}, Hg1212 in~\cite{ItohHg1212}, Bi$_{2.1}$Sr$_{1.9}$Ca$_{0.9}$Cu$_2$O$_{8+\delta}$ in~\cite{Nishiyama}, YBa$_2$Cu$_3$O$_{6.93}$ in~\cite{ItohZnYBCO} and YBa$_2$Cu$_4$O$_8$ in~\cite{ItohZnYBCO}. 
Solid curves are simulations using multiple Gaussian functions.
More than two pairs of $^{63}$Cu and $^{65}$Cu NQR lines were needed to reproduce
the broad NQR spectra except YBa$_2$Cu$_3$O$_7$ and YBa$_2$Cu$_4$O$_8$. 

Non-stoichiometry in compounds gives rise to inhomogeneous broadening in NMR and NQR spectra,
because magnetic shift and quadrupole frequency are distributed. 
YBa$_2$Cu$_4$O$_8$ with double CuO$_{2}$ planes and double CuO chains in unit cell
has $T_\mathrm{c}$ = 82 K.
In Fig.~\ref{CuNQR}, 
this stoichiometric and naturally underdoped compound shows sharp Cu NQR spectra. 
YBa$_2$Cu$_3$O$_{6.93}$ with double CuO$_{2}$ planes and a single CuO chain in unit cell
has $T_\mathrm{c}$ = 93 K.
In Fig.~\ref{CuNQR}, 
this stoichiometric and slightly overdoped compound shows sharp Cu NQR spectra.  
Although the CuO$_2$ planes of these compounds have the buckling structures,  
the Cu NQR spectra are rather sharp.  

The superconducting Hg1201 and Hg1212 at any doping level,
however, show the inhomogeneously broad Cu NQR spectra~\cite{GippiusHg1201,OhsugiHg,JulienHg1212}.
In general, crystalline imperfection is the origin of the inhomogeneous broadening of an NQR spectrum.
In Fig.~\ref{CuNQR}, Bi$_{2.1}$Sr$_{1.9}$Ca$_{0.9}$Cu$_2$O$_{8+\delta}$ (Bi2212) also shows a broad Cu NQR spectrum.
This is attributed to a wide range structural modulation of the BiO layers.
In spite of  the dilute oxygen concentration in the HgO layers of Hg1201 and Hg1212,
most of the Cu nuclei feel inhomogeneous electric field gradients. 
Thus, the effect of dilute oxygen ions on the crystalline potential may be long ranged. 
Long-range Friedel oscillations from the excess oxygen ions may yield such a broad NQR spectrum.
The inhomogeneously broad Cu NQR spectrum is observed in the nearly perfect flat CuO$_2$ plane.
This is instructive for us to understand the relation between NQR and the crystal structure. 
The broad Cu NQR spectrum results from the nonstoichiometry but not from the buckling of the CuO$_2$ planes.  

In Fig.~\ref{CuNQR}, one should note that the value of Cu $\nu_\mathrm{NQR}$ of Hg1201 is similar to that of Tl$_2$Ba$_2$CuO$_{6+\delta}$ and about a half of those of YBa$_2$Cu$_3$O$_7$ and YBa$_2$Cu$_4$O$_8$. The local density approximation calculations of the NQR frequencies account for these experimental similarity and difference~\cite{Singh}. 

\subsection{Bulk magnetic susceptibility}
A static magnetic field $H$ is applied to a material along the $z$-axis and then
a magnetization $M_z$ is measured by a magnetometer. 
The bulk magnetization $M_z$ is the sum of respective electron spins ${\langle S_{iz}\rangle}$  
\begin{equation}
M_z = \sum_{i}{\langle S_{iz}\rangle},
\label{bulkM}
\end{equation}
where $\langle S_{i\alpha}\rangle$ is a thermal average of the electron spin along the $\alpha$-axis.
Then the bulk magnetic susceptibility $\chi$ is defined by 
\begin{equation}
\chi = {\sum_{i}{\langle S_{iz}\rangle} \over H_\mathrm{ext}}.
\label{Xs}
\end{equation}
The paramagnetism of 3$d$ transition metal oxides results from
the unpaired $d$ electron spins and the orbital momentums.
The bulk magnetic susceptibility is  expressed by
the sum of the spin susceptibility $\chi_\mathrm{spin}$, the Van Vleck orbital susceptibility $\chi_\mathrm{vv}$, and the diamagnetic susceptibility $\chi_\mathrm{dia}$ of inner core electrons 
\begin{equation}
\chi=\chi_\mathrm{spin}(T)+\chi_\mathrm{vv}+\chi_\mathrm{dia}.
\label{bulkX}
\end{equation}
\begin{figure}
\begin{center}
\includegraphics[width=0.8\linewidth]{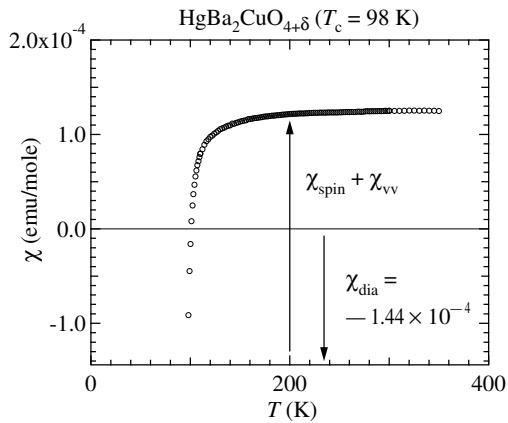}
\end{center}
\caption{
Uniform magnetic susceptibility $\chi$ of the optimally doped powdered sample of Hg1201 ($T_\mathrm{c}$ = 98 K) in~\cite{YamamotoHg1201} at a magnetic field of 1 Tesla~\cite{ItohXT}. 
}
\label{XTHg1201}
\end{figure}  

Fgiure~\ref{XTHg1201} shows the uniform magnetic susceptibility $\chi$ of the optimally doped powdered sample of Hg1201 ($T_\mathrm{c}$ = 98 K) in~\cite{YamamotoHg1201} at a magnetic field of 1 Tesla~\cite{ItohXT}. The pseudo spin-gap behavior is seen in the bulk magnetic susceptibility.
The magnitude of  $\chi$ above 150 K is the same order of the other high-$T_\mathrm{c}$ superconducting cuprates .

\subsection{Knight shift}
Nuclear spins of constituent ions in a crystal interact with electron spins through an electron-nuclear hyperfine coupling. 
When a static uniform  magnetic field $H_\mathrm{ext}$ is applied to the system along the $z$-axis, 
the electron medium is polarized along the $z$-axis. 
The magnetic polarization produces an additional magnetic field at the nuclear sites through the hyperfine coupling 
and then leads to a shift of the resonance field (frequency) of the nuclear spins. 
This is Knight shift $K$. 

The static nuclear spin Hamiltonian of a Zeeman coupling and the electron-nuclear hyperfine coupling is given by 
\begin{equation}
H_\mathrm{int} = -\gamma_n I\cdot \sum_{i}{(H_\mathrm{ext}+A_{iz}\langle S_{iz}\rangle)}
\label{HF}
\end{equation}
where $\gamma_n$ is the nuclear gyromagnetic ratio and $A_{i\alpha}$ is a hyperfine coupling constant along the $\alpha$-axis with the $i$-site electron spin.

The experimental Knight shift is defined by a shift of the observed resonance frequency $\omega_\mathrm{res}$
from the reference frequency $\omega_\mathrm{ref}$
\begin{eqnarray}
\omega_\mathrm{res} &=&
\gamma_n H_\mathrm{ext}(1+K) 
\nonumber \\ &=&
\omega_\mathrm{ref}(1+K).
\label{Knight}
\end{eqnarray}

In the 3$d$ transition metal oxides, the Knight shift $K$ is decomposed into 
the spin shift $K_\mathrm{s}$ due to unpaired electron spins and the orbital shift $K_\mathrm{orb}$ due to the Van Vleck orbital susceptibility
\begin{equation}
K = K_\mathrm{s}(T)+K_\mathrm{orb}.
\label{KsKvv}
\end{equation} 
The spin Knight shift $K_{\alpha}$ is given by 
\begin{equation}
K_{s} = A(q = 0){\chi_{s}^{'}(q = 0, \omega = 0) \over N_\mathrm{A}\mu_\mathrm{B}}.
\label{Ks}
\end{equation}
where $A(q = 0)$ is the uniform Fourier component of the hyperfine coupling constant 
and $\chi\prime (q = 0, \omega=0)$ is the static ($\omega$ = 0) uniform ($q$ = 0) electron spin susceptibility.   
$\mu_\mathrm{B}$ is the Bohr magneton.  
$N_\mathrm{A}$ is the Avogadro number. 
$\chi_{\alpha}\prime$ is the measured bulk susceptibility in emu/mole-atom. 

The hyperfine coupling constant reflects the characters of the wave functions of the electron orbitals. 
Measurement of the Knight shift at each nuclear enables us to obtain the site-specific information. 
The temperature dependence of the Knight shift reveals that of the intrinsic uniform spin susceptibility $\chi_\mathrm{spin}$ at each site.  

In covalent bonded compounds, the transferred and supertransferred hyperfine coupling constants play important roles. 
In the high-$T_\mathrm{c}$ cuprate superconductors, 
the Fourier transformed hyperfine coupling constants at the plane-site Cu and the oxygen are expressed by 
\begin{equation}
A_\alpha^{Cu}({\bf q}) = A_\alpha+2B\{ \mathrm{cos}(q_x)+\mathrm{cos}(q_y)\} 
\label{CuAq}
\end{equation} 
and 
\begin{equation}
A_\alpha^{O}({\bf q}) = 2C\mathrm{cos}(q_x/2), 
\label{OAq}
\end{equation} 
where $A_\alpha$ ($\alpha$ = $ab$ and $cc$) is an on-site anisotropic hyperfine coupling constant,
$B$($>$ 0) is a supertransferred isotropic hyperfine coupling constant from a Cu to a Cu through an oxygen,
and  $C$($>$ 0) is a transferred hyperfine coupling constant from a Cu to an oxygen~\cite{ZR,Russ}.  

For the optimally doped Hg1201, we estimate
$A_{ab}^{Cu}({\bf q}=0)$ = 145 kOe/mole-Cu-$\mu_\mathrm{B}$ from the $K-\chi$ plot,
where the Knight shift $K_{ab}$~\cite{ItohHg1201F} is plotted against the bulk magnetic susceptibility $\chi$ in Fig.~\ref{XTHg1201}
with temperature as an implicit parameter. 

\subsection{Nuclear spin-lattice relaxation}
Nuclear spins are coupled by the fluctuating hyperfine fields of electron spins
through a time-dependent hyperfine coupling Hamiltonian
\begin{equation}
H_\mathrm{int}(t)={1 \over 2}\sum_{i}{(I_{-}A_{i}{\delta}S_{i+}(t)+I_{+}A_{i}{\delta}S_{i-}(t))}.
\label{HF}
\end{equation}
In a spin-echo recovery technique, the nuclear moments excited by an inversion $rf$-pulse 
are in a thermal non-equilibrium state. The energy dissipation from the nuclear moments to a lattice 
takes place through the fluctuating hyperfine fields. The recovery time of the nuclear moments
to a thermal equilibrium state is the nuclear spin-lattice relaxation time $T_\mathrm{1}$. 

Moriya derived a general expression of  $T_\mathrm{1}$~\cite{MoriyaT1},
\begin{eqnarray}
{1 \over T_1}&=&
{\gamma_n^2 \over 2}\sum_{i}{A_{i}^{2}\int_{-\infty}^{\infty}{\langle\{{\delta}S_{i+}(t){\delta}S_{i-}(0)\}\rangle\mathrm{e}^{-i{\omega_n}t}\mathrm{d}t}}
\nonumber \\ &=&
{2\gamma_n^2 \over g^2\mu_\mathrm{B}^2}{k_\mathrm{B}T \over \omega_n}\int{\mathrm{d}^DqA_{ab}(q)^{2}\chi_{ab}{}^{''}(q, \omega_n)}.
\label{T1}
\end{eqnarray} 
$D$ is the dimension of space on which the electronic system lies. 
1/$T_{1}T$ is the square of the hyperfine coupling constant $A(q)^2$ times the wave vector averaged low frequency dynamical spin susceptibility $\chi^{''}({\bf q}, \omega)$~\cite{MoriyaT1}. 
Thus, from measurements of the Knight shift and the nuclear spin-lattice relaxation time,
one can infer the $q$ dependence of the dynamical spin susceptibility $\chi "({\bf q}, \omega)$.
The modified Korringa ratio can  give us a criterion which the electronic system is, ferromagnetic or antiferromagnetic and where $\chi^{''}({\bf q}, \omega)$ is enhanced in the  $q$ space. 

In general, the $\bf q$ dependence of the hyperfine coupling constant $A({\bf q})$ is slower than that of $\chi({\bf q})$. 
But, the form factor of the plane-site oxygen $A_\alpha^{O}({\bf q})^2$ of Eq.~(\ref{OAq}) acts as a filter for a finite $\bf q$ correlation.  
The antiferromagnetic correlation of $\bf Q$ = [$\pi$, $\pi$] and $\bf Q^{*}$ = [$\pi$(1$\pm\delta$), $\pi$(1$\pm\delta$)] 
is cancelled at  the plane-site oxygen through the form factor of Eq.~(\ref{OAq}). 
The plane-site Cu nuclei can probe the antiferromagnetic correlation through Eq.~(\ref{CuAq}) . 

\subsection{Gaussian Cu nuclear spin-spin relaxation}  
Strong indirect Cu nuclear spin-spin interaction was first found in YBa$_2$Cu$_3$O$_7$~\cite{PS1,PS2}. 
Gaussian decay in the transverse relaxation of the plane-site Cu nuclear moments
is predominantly induced by the indirect nuclear spin-spin interaction through the in-plane 
antiferromagnetic electron spin susceptibility.
Strong antiferromagnetic fluctuations persist 
in all the high-$T_\mathrm{c}$ cuprate superconductors. 
Thus, the Gaussian decay rate of the plane-site Cu nuclear spin-spin relaxation 
provides us fruitful information on 
the antiferromagnetic electron spin susceptibility~\cite{PS2,ItohT2g1,ItohT2g2,ItohT2g3,ItohT2g4,ItohT2g5}.    
  
The indirect nuclear spin-spin interaction is given by 
\begin{equation}
H_\mathrm{II} = \sum_{i, j}{\Phi(r_{ij})I_{j}\cdot I_{j}}  
\label{IIcoupling}
\end{equation}
where a range function $\Phi$(r$_{ij}$) is given by  
\begin{equation}
\Phi(r_{ij}) = \int{\mathrm{d}^DqA(q)^2\chi^{'}(q)}. 
\label{RangeFunc}
\end{equation}
The Gaussian decay rate 1/$T_{2g}$ of the nuclear spin-spin relaxation is given by
\begin{eqnarray}
\left({1 \over T_{2g}}\right)^2&\propto&
\sum_{j}{\Phi(r_{ij})^2}
\nonumber \\ &\approx&  
\int{\mathrm{d}^DqA(q)^4\chi^{'}(q)^2}. 
\label{range}
\end{eqnarray}

The Kramers-Kronig relation is  
\begin{equation}
\chi^{'}(q) = {2 \over \pi}\int_{0}^{\infty}{\mathrm{d}\omega {\chi^{''}(q, \omega) \over \omega}}. 
\label{KK}
\end{equation}
The Gaussian decay rate 1/$T_{2g}$ reflects the full range frequency integration 
of the dynamical spin susceptibility $\chi^{''}$($\bf\it q$, $\omega$).

The cross section of inelastic neutron scattering is expressed by
\begin{equation}
{{\partial^2 \sigma} \over {\partial\omega\partial\Omega}} \propto {1 \over {1-e^{-\beta\omega}}} \chi^{''}(q, \omega),
\label{INS}
\end{equation}
where $q$ is a momentum transfer of the neutron, $\omega$ the energy transfer, and  $\beta$ = 1/$k_\mathrm{B}T$~\cite{Marshall}. 
In the high-$T_\mathrm{c}$ cuprate superconductors, 
the enhancement of $\chi^{''}$($\bf\it q$, $\omega$) was observed over a finite range of $\omega$ and at around $\it q$ = $\bf Q$ or $\bf Q^*$~\cite{RM}. 
In general, it is hard to see $\chi^{''}$($\bf\it q$, $\omega$) over the full range.    

From measurements of the nuclear spin-lattice relaxation and the Gaussian decay rate of nuclear spin-spin relaxation, 
one can infer the dynamical spin susceptibility at low and high frequency regions. 

\subsection{Two dimensional nearly antiferromagnetic spin fluctuation model}
The electronic state of a CuO$_2$ plane of the high-$T_\mathrm{c}$ cuprate superconductor
is described by a single band picture. 
The anisotropy of Knight shift and nuclear spin-lattice relaxation and the site differentiations 
are explained by anisotropic hyperfine coupling and different $\bf q$ dependence of 
the coupling constant due to transferred and supertransferred hyperfine couplings. 
That is Mila-Rice-Shastry hyperfine coupling Hamiltonian~\cite{MilaRice,Shastry}. 
Two dimensional antiferromagnetic spin fluctuations through the Mila-Rice-Shastry hyperfine coupling yield the anisotropy and the site difference~\cite{Imai,Takigawa,Walstedt}.
The two dimensional spin fluctuation models were successfully applied 
to account for NMR, neutron scattering and conductivity results~\cite{MMP,MTU,RPA}.  
 
We employ the two dimensional nearly antiferromagnetic spin fluctuation model in~\cite{MTU}.
The dynamical spin susceptibility 
is expressed by a relaxation mode
\begin{equation}
\chi (q, \omega) = {\chi (q) \over {1-i \omega / \Gamma (q)}}.  
\label{Xqw}
\end{equation}
The leading terms of $\chi(q)$ and $\Gamma(q)$ in a random phase approximation (RPA) are expressed using a long wave length expansion around a specific mode $q$ = $Q$ by
\begin{equation}
\chi (q) \approx {\chi_0(Q) \over {\kappa ^2+(q-Q)^2}}  
\label{Xq}
\end{equation}
and
\begin{equation}
\Gamma (q) \approx {\Gamma_0(Q) (\kappa ^2+(q-Q)^2)}.  
\label{Gq}
\end{equation}
$\kappa$ is the inverse of a magnetic correlation length $\xi$ defined around $q$ = $Q$,
\begin{equation}
\xi^2 = - {1 \over \chi(Q)}{\partial^2 \chi(q) \over \partial q^2}\Bigr|_{q=Q}.  
\label{Gq}
\end{equation}
$\chi_0(Q)$ is the spin fluctuation amplitude at $q$ = $Q$ and 
$\Gamma_0(Q)$ is the characteristic spin fluctuation energy. 
The staggered spin susceptibility $\chi({\bf Q})$ is then 
 \begin{equation}
\chi ({\bf Q}) = \chi_0({\bf Q})\xi^2.  
\label{XsQ}
\end{equation}
 
We obtain the plane-site Cu nuclear spin-lattice relaxation rate expressed by the leading term of $\xi$
 \begin{eqnarray}
{1 \over T_{1}T} &\approx&
 A(Q)^2\int{\mathrm{d}^Dq {\chi(q) \over \Gamma (q)}}  
\nonumber \\ &\propto&
{\chi_0(Q) \over \Gamma_0(Q)}\int{\mathrm{d}^Dq \over {\{\kappa^2+(q-Q)^2\}^2}}
\nonumber \\ &\propto&
\xi^{4-D} {\chi_0(Q) \over \Gamma_0(Q)}\int_{0}^{\xi q_\mathrm{B}}{\mathrm{d}^D(\xi \tilde{q}) \over \{{1+(\xi \tilde{q})^2\}^2}}
\nonumber \\ &\propto&
\xi^{4-D}
\nonumber \\ &\propto&
\chi(Q)^{2-D/2}, 
\label{T1TRPA}
\end{eqnarray}
where $q_\mathrm{B}$ is the spherical radius of the same volume as the first Brillouin zone and $\xi q_\mathrm{B}\gg$ 1. 
We also obtain the Gaussian Cu nuclear spin-spin relaxation rate expressed by the leading term of $\xi$
 \begin{eqnarray}
\left({1 \over T_{2g}}\right)^2&\approx&
 A(Q)^4\int{\mathrm{d}^Dq {\chi(q)^2}}  
\nonumber \\ &\propto&
{\chi_0(Q)}^2 \int{\mathrm{d}^Dq \over {\{\kappa^2+(q-Q)^2\}^2}}
\nonumber \\ &\propto&
\xi^{4-D} \chi_0(Q)^2 \int{\mathrm{d}^D(\xi \tilde{q}) \over {\{1+(\xi \tilde{q})^2\}^2}}
\nonumber \\ &\propto&
\xi^{4-D}
\nonumber \\ &\propto&
\chi(Q)^{2-D/2}. 
\label{T2gRPA}
\end{eqnarray}
Thus, for $D$ = 2 we have
\begin{equation}
{T_1T \over {T_{2g}}} \propto \Gamma_0(Q) \xi^{-1},    
\label{Z1}
\end{equation}
and
\begin{equation}
{T_1T \over {(T_{2g})^2}} \propto \Gamma_0(Q)\chi_0(Q).    
\label{Z2}
\end{equation}

In the self-consistent renormalization (SCR) theory for two dimensional antiferromagnetic spin fluctuations, 
the Curie-Weiss behavior of the sqaure of the antiferromagnetic correlation length $\xi^2$ ($\propto$ the staggered spin susceptibility $\chi(Q)$)
is reproduced as a function of the distance from the quantum critical point and the spin fluctuation energy $\Gamma_0(Q)$~\cite{MTU}. 
Using $\chi_0(Q)$ = 1/2$\alpha_{s}T_\mathrm{A}$ ($\alpha$ = $U\chi_{0}(Q)$), $\Gamma_0(Q)$ = 2$\pi T_\mathrm{0}$ and $t$ = $T/T_0$, 
we obtain~\cite{ItohHg1212} 
\begin{equation}
{T_1T \over {T_{2g}}} \propto T_0 \xi(t)^{-1},    
\label{SCR1}
\end{equation}
which represents the inverse of the antiferromagnetic correlation length with a unique parameter $T_0$,
and
\begin{equation}
{T_1T \over {(T_{2g})^2}} \propto {T_0 \over {T_A}},    
\label{SCR2}
\end{equation}  
which represents the integrated spin fluctuation weight. 
Using Eqs.~(\ref{SCR1}) and~(\ref{SCR2}), 
the NMR relaxation data for Tl$_2$Ba$_2$CuO$_{6+\delta}$,
YBa$_2$Cu$_3$O$_7$ and YBa$_2$Cu$_4$O$_8$ were analyzed~\cite{ItohT2g2,ItohT2g3,ItohT2g4,ItohT2g5}. 

\section{NORMAL-STATE PSEUDO SPIN-GAP}
\subsection{Single-layer pseudo spin-gap: Hg1201} 
\begin{figure}
\begin{center}
\includegraphics[width=1.1\linewidth]{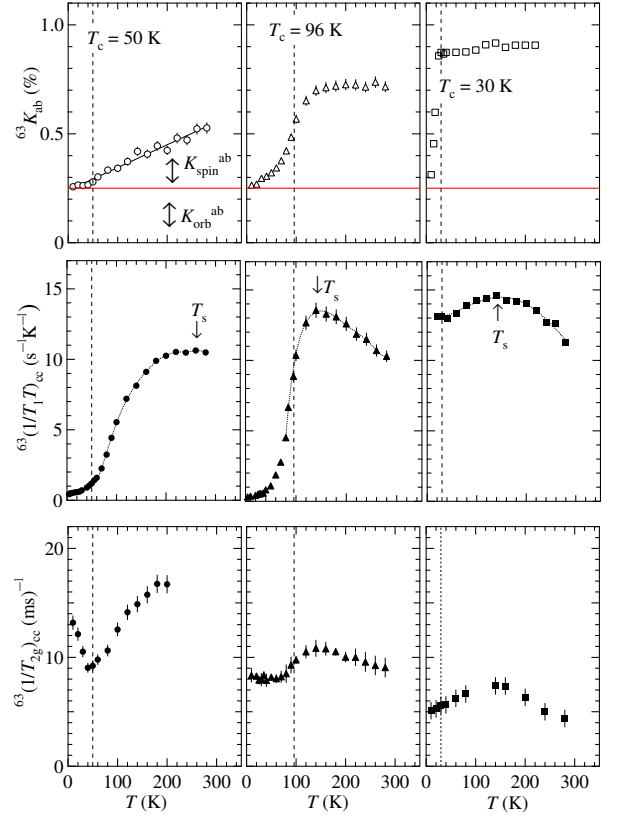}
\end{center}
\caption{
$^{63}$Cu NMR results of single-layer superconductors Hg1201. 
Temperature dependences of the plane-site $^{63}$Cu Knight shifts  $K_\mathrm{ab}$ (top panels),  nuclear spin-lattice relaxation rates (1/$T_1T)_\mathrm{cc}$ (middle panels), and Gaussian decay rates (1/$T_\mathrm{2g})_\mathrm{cc}$ of nuclear spin-spin relaxation (bottom panels) are
shown for the underdoped (left), the optimally doped (center) and the overdoped samples (right).
Dash lines denote the respective $T_\mathrm{c}$'s. 
The data are reproduced from~\cite{ItohHg1201L,ItohHg1201F,ItohHg1201P}.
}
\label{fig:Hg1201F1}
\end{figure} 
The left panels of Fig.~\ref{fig:Hg1201F1} show $^{63}$Cu NMR results of the underdoped Hg1201 of $T_\mathrm{c}$ = 50 K~\cite{ItohHg1201L,ItohHg1201F}. 
The NMR experiments were performed for the powdered polycrystalline sample. 
All the powder samples were magnetically aligned along the $c$ axis. 
In general the mercury compounds cannot be easily aligned by a magnetic field. 
Then, the NMR experiments were performed for partially oriented powder samples.  
This does not indicate that the NMR data are the partially powder-averaged ones. 
The sharply aligned NMR lines can be separated in the powder pattern,
so that the selected signals surely come from the aligned grains. 

The plane-site $^{63}$Cu Knight shift $K_\mathrm{ab}$,  nuclear spin-lattice relaxation rate (1/$T_1T)_\mathrm{cc}$, and Gaussian nuclear spin-echo decay rate (1/$T_\mathrm{2g})_\mathrm{cc}$ are shown as functions of temperature. 
The subscripts of $cc$ and $ab$ denote the data in the external magnetic field $H_\mathrm{ext}\sim$ 8 Tesla along the $c$ and $ab$ axis, respectively.   
No appreciable field dependence was observed within $H_\mathrm{ext}$ = 4 $\sim$ 8 Tesla.

The drastic decreases of the Cu Knight shift $K$ and nuclear spin-lattice relaxation rate 1/$T_1T$ with cooling down are clearly seen from room temperature. 
This is an obvious evidence of the existence of the pseudo spin-gap in the single layer cuprate.
The decrease of the Cu Knight shift is independently found in~\cite{Hoffmann}. 
The pseudo spin-gap behavior of the uniform spin susceptibility is also found by the in-plnae $^{17}$O NMR experiments~\cite{BobroffHg1201}. 
In passing, for the underdoped triple-layer Hg1223,
the pseudo spin-gap behavior has been observed by Cu NMR experiments~\cite{JulienHg1223}. 
The absence of the Hebel-Slichter peak of 1/$T_1T$ just below $T_\mathrm{c}$ 
excludes the weak coupling $s$-wave pairing symmetry. 

The slow decrease of the Cu 1/$T_\mathrm{2g}$ is also seen below 200 K. 
It suggests that the large pseudo spin-gap leads to the loss of the total weight 
of the frequency integrated $\chi^{"}(q, \omega)$ and then that the static staggered spin susceptibility $\chi^{'}(\bf Q)$ decreases.
The drastic decrease of 1/$T_1T$ but the moderate decrease of 1/$T_\mathrm{2g}$ are
theoretically reproduced by the numerical calculations involving the self-energy correction due to the enhanced $d_{x^2-y^2}$-wave superconducting fluctuations~\cite{Yanase}.
In this theory, the pseudogap is a consequence from the resonance scattering in the two dimensional strong coupling superconductivity.  
The finite 1/$T_\mathrm{2g}$ below $T_\mathrm{c}$ is an evidence for the $d_{x^2-y^2}$ wave pairing symmetry~\cite{BulutScalapino,ItohYoshimura,ItohD}.  

Figure~\ref{fig:Hg1201F1} shows  
the hole doping dependence of $^{63}$Cu NMR results of Hg1201~\cite{ItohHg1201L,ItohHg1201F,ItohHg1201P}. 
Temperature dependences of the plane-site $^{63}$Cu Knight shifts  $K_\mathrm{ab}$ (top panels),  nuclear spin-lattice relaxation rates (1/$T_1T)_\mathrm{cc}$ (middle panels), and Gaussian decay rates (1/$T_\mathrm{2g})_\mathrm{cc}$ of nuclear spin-spin relaxation (bottom panels) are
shown for the underdoped (left), the optimally doped (center) and the overdoped samples (right).
The pseudo spin-gap temperature $T_\mathrm{s}$ is defined by the maximum temperature or the onset of the decrease of (1/$T_1T)_\mathrm{cc}$. 
$T_\mathrm{s}$ decreases from the underdoped to the overdoped samples. 
The temperature region of the Curie-Weiss behavior of (1/$T_1T)_\mathrm{cc}$ above $T_\mathrm{s}$ 
is broadened with the hole doping.  

\subsection{Double-layer pseudo spin-gap: Hg1212}
\begin{figure}
\begin{center}
\includegraphics[width=1.0\linewidth]{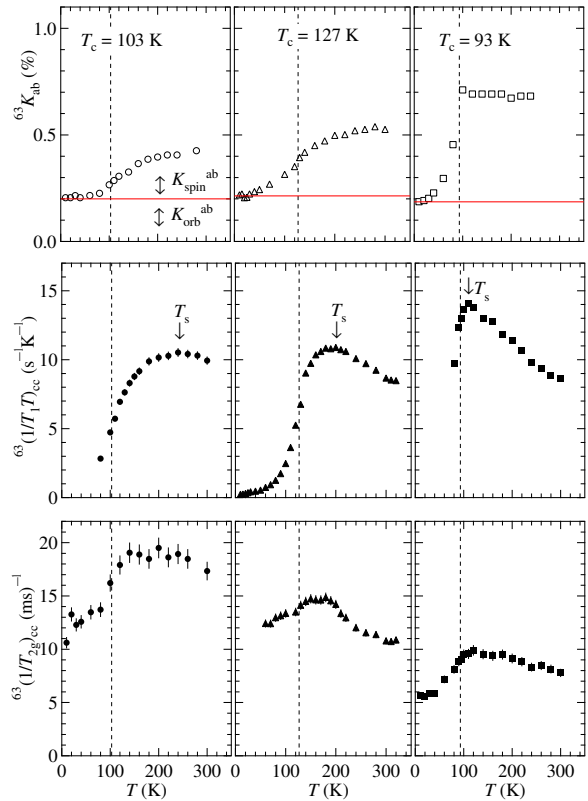}
\end{center}
\caption{
$^{63}$Cu NMR results of double-layer superconductors Hg1212. 
Temperature dependences of the plane-site $^{63}$Cu Knight shifts  $K_\mathrm{ab}$ (top panels),  nuclear spin-lattice relaxation rates (1/$T_1T)_\mathrm{cc}$ (middle panels), and Gaussian decay rates (1/$T_\mathrm{2g})_\mathrm{cc}$ of nuclear spin-spin relaxation (bottom panels) are
shown for the underdoped (left), the optimally doped (center) and the overdoped samples (right).
Dash lines denote the respective $T_\mathrm{c}$'s. 
The data are reproduced from~\cite{ItohHg1212}.
}
\label{fig:Hg1212F1}
\end{figure}  
The middle panels of Fig.~\ref{fig:Hg1212F1} show $^{63}$Cu NMR results of the optimally doped Hg1212 of $T_\mathrm{c}$ = 127 K~\cite{ItohHg1212},
which is the maximum  $T_\mathrm{c}$ among the ever reported single and double layers. 
The plane-site $^{63}$Cu Knight shift $K_\mathrm{ab}$,  nuclear spin-lattice relaxation rate (1/$T_1T)_\mathrm{cc}$, and Gaussian nuclear spin-echo decay rate 1/$T_\mathrm{2g}$ are shown as functions of temperature.   
The decrease of the $^{63}$Cu Knight shift $K_\mathrm{ab}$ with cooling down is obvious even at the optimal Hg1212. 
Above $T_\mathrm{s}$ = 200 K, 1/$T_1T$ shows a Curie-Weiss behavior, 
because of the development of the antiferromagnetic correlation length. 
The decrease of 1/$T_1T$ starts below about $T_\mathrm{s}$ = 200 K.  
This $T_\mathrm{s}$ = 200 K of Hg1212  is larger than $T_\mathrm{s}$ = 140 K  of the optimally doped Bi2212 with $T_\mathrm{c}$ = 86 K in~\cite{Ishida}. 
The record optimal $T_\mathrm{c}$ of Bi2212 is about 96 K~\cite{Kato}.   
For the double layer systems of Hg1212 and Bi2212,
the ratio ($\sim$ 1.4) of the optimal $T_\mathrm{s}$'s is nearly the same as that of $T_\mathrm{c}$.   
The slight decrease of 1/$T_\mathrm{2g}$ is also seen, 
because of the large pseudo spin-gap effect on the frequency integrated $\chi^{"}(q, \omega)$. 
The finite 1/$T_\mathrm{2g}$  below $T_\mathrm{c}$ indicates the $d_{x^2-y^2}$ wave pairing symmetry~\cite{BulutScalapino,ItohYoshimura,ItohD}.   

Figure~\ref{fig:Hg1212F1} shows  
the hole doping dependence of $^{63}$Cu NMR results of Hg1212~\cite{ItohHg1212}. 
Temperature dependences of the plane-site $^{63}$Cu Knight shifts  $K_\mathrm{ab}$ (top panels),  nuclear spin-lattice relaxation rates (1/$T_1T)_\mathrm{cc}$ (middle panels), and Gaussian decay rates (1/$T_\mathrm{2g})_\mathrm{cc}$ of nuclear spin-spin relaxation (bottom panels) are
shown for the underdoped (left), the optimally doped (center) and the overdoped samples (right). 
The pseudo spin-gap temperature $T_\mathrm{s}$ decreases from the underdoped to the overdoped samples. The values of $T_\mathrm{s}$  of Hg1212 are higher than those of Bi2212~\cite{Ishida}. 
The temperature region of the Curie-Weiss behavior in (1/$T_1T)_\mathrm{cc}$ above $T_\mathrm{s}$ 
is broadened with the hole doping. 
It should be noted that the overdoped sample shows the Curie-Weiss behavior in (1/$T_1T)_\mathrm{cc}$ but not Korringa behavior above $T_\mathrm{s}$~\cite{JulienHg1212}. 

\subsection{Interlayer coupling via Hg NMR} 
\begin{figure}
\begin{center}
\includegraphics[width=0.7\linewidth]{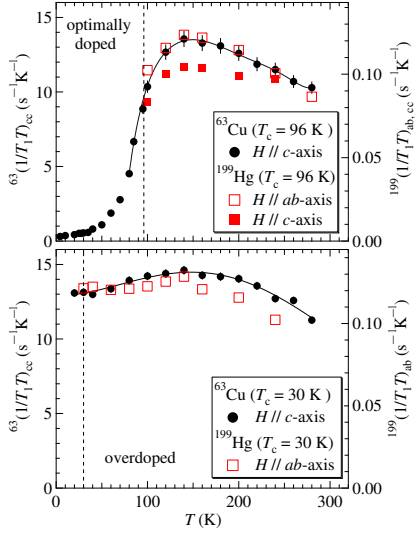}
\end{center}
\caption{
$^{63}$Cu and $^{199}$Hg nuclear spin-lattice relaxation rates 1/$T_1T$ for optimally doped Hg1201 (a) 
and for overdoped Hg1201 (b)~\cite{ItohHg1201F}.  
Temperature dependence of $^{199}(1/T_1T)$ is nearly the same as that of $^{63}(1/T_1T)_{cc}$
for both samples.   
}
\label{CuvsHgT1}
\end{figure} 
The Hg site is located just halfway between the CuO$_2$ planes
and between the respective Cu ions. The Hg nuclei can serve as a probe
of the interplane coupling. If the interplane coupling is antiferromagnetic,  
the in-plane antiferromagnetic correlation is also cancelled out at the Hg site.
If the inteplane coupling is uniform and ferromagnetic, 
the in-plane antiferromagnetic correlation is also seen at the Hg site. 

Figure~\ref{CuvsHgT1} shows $^{63}$Cu and $^{199}$Hg nuclear spin-lattice relaxation rates 1/$T_1T$ for the optimally doped Hg1201 (a) and for the overdoped Hg1201 (b)~\cite{ItohHg1201F}.  
The temperature dependence of $^{199}(1/T_1T)$ is nearly the same as that of $^{63}(1/T_1T)$
for both samples.
Thus, the inteplane coupling is not antiferromagnetic but uniform. 
This uniform interplane coupling is also observed by  $^{63}$Cu and $^{199}$Hg NMR 1/$T_1T$ for Hg1212~\cite{JulienHg1212,HorvaticHg}. 
In contrast to the reports ~\cite{Suh1,Suh2} that $^{63}$Cu and $^{199}$Hg nuclear spin-lattice relaxation rates 1/$T_1T$ show the different behaviors with each other in Hg1201,  
these results in Fig.~\ref{CuvsHgT1} indicate that the interplane coupling is uniform in Hg1201 irrespective of the doping level. 

\subsection{Pseudo spin-gap phase diagram}                  
\begin{figure}
\begin{center}
\includegraphics[width=0.8\linewidth]{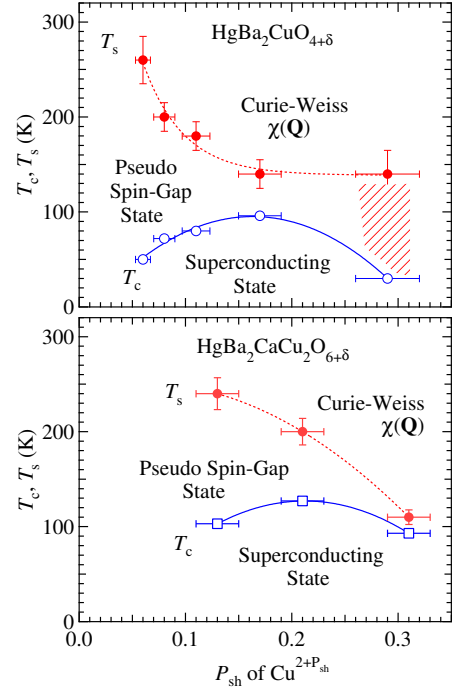}
\end{center}
\caption{
Magnetic phase diagrams of Hg1201 and Hg1212~\cite{ItohHg1201F,ItohHg1212,ItohHg1201P}. 
Pseudo spin-gap temperatures $T_\mathrm{s}$ and superconducting transition temperatures $T_\mathrm{c}$ are plotted against the hole concentration $P_\mathrm{sh}$ per plane-site Cu.
The value $P_\mathrm{sh}$ of Cu$^{2+P_\mathrm{sh}}$ is estimated from the excess oxygen concentration $\delta$ and the charge neutrality condition. 
Solid and dash curves are guides to the eyes. 
}
\label{MPhase}
\end{figure} 
Figure~\ref{MPhase} shows the magnetic phase diagrams of Hg1201(a) and Hg1212 (b)~\cite{ItohHg1201F,ItohHg1212,ItohHg1201P}. 
Pseudo spin-gap temperatures $T_\mathrm{s}$ and superconducting transition temperatures $T_\mathrm{c}$ are plotted against the hole concentration $P_\mathrm{sh}$.  
The hatched region of the overdoped Hg1201 indicates a pseudo Korringa behavior 
below $T_\mathrm{s}$. 
It should be noted that the pseudo spin-gap persists at the optimally doped regions for Hg1201 and H1212. 
The doping dependence of $T_\mathrm{s}$ of Hg1212 is different from that of Hg1201
even in the underdoped region. The $T_\mathrm{s}$ as a function of $P_\mathrm{sh}$
does not seem to be universal. 

\begin{figure}
\begin{center}
\includegraphics[width=0.8\linewidth]{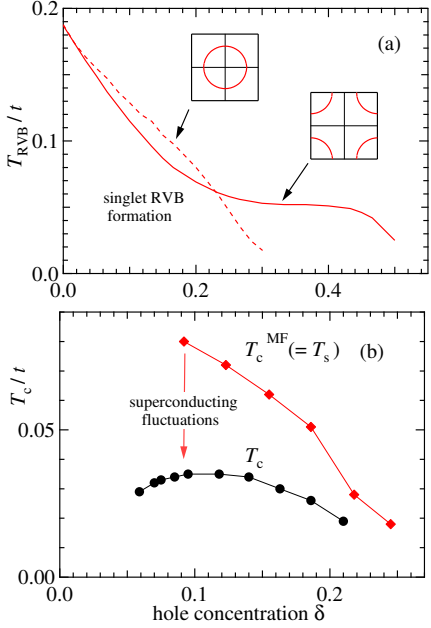}
\end{center}
\caption{
(a) Theoretical hole doping dependences of pseudo spin-gap temperatures $T_\mathrm{s}$ 
as siglet-RVB formation temperatures $T_\mathrm{RVB}$ in a two dimensional $t$-$J$
model with two type Fermi surfaces reproduced from~\cite{TKF}.
$t$ is a Cu-to-Cu transfer integral and $J$ is a superexchange interaction in charge-transfer type compounds. 
The inset figures are the two type Fermi surfaces. 
(b) Theoretical hole doping dependences of the mean-field $T_\mathrm{c}^\mathrm{\mathrm{MF}}$ and the true $T_\mathrm{c}$ suppressed by $d_{x^2-y^2}$-wave superconducting fluctuations reproduced from~\cite{Yanase}.
}
\label{theorySG}
\end{figure} 
We present two theoretical explanations for the doping dependence of $T_\mathrm{s}$.
In the two dimensional $t$-$J$ model with spinon-holon decomposition technique,
the pseudo spin-gap temperature $T_\mathrm{s}$ is regarded as the onset temperature $T_\mathrm{RVB}$ of a spinon singlet RVB (resonating valence bond) state~\cite{Kotliar,Suzumura,TKF}. 
The real transition $T_\mathrm{c}$ is given by a Bose-Einstein condensation temperature $T_\mathrm{BEC}$ of holons, leading to the underdoped regime. 
With an existing approximation, $T_\mathrm{RVB}$ is a second order phase transition temperature. 
Figure~\ref{theorySG} (a) shows a numerical $T_\mathrm{RVB}$ as a function of
doped hole concentration~\cite{TKF}. 
The doping dependence of $T_\mathrm{RVB}$ depends on the contour of a basal Fermi surface. 
For each high-$T_\mathrm{c}$ family with different Fermi surface, $T_\mathrm{RVB}$ exhibits the different doping dependence. 
 
In the two dimensional superconducting fluctuation theory with the strong coupling,   
the pseudo spin-gap temperature $T_\mathrm{s}$ is regarded as the onset of enhancement
of $d_{x^2-y^2}$-wave superconducting fluctuations and a mean-field $T_\mathrm{c}$. 
The actual $T_\mathrm{c}$ is reduced by the strong superconducting fluctuations
so that the underdoped regime appears.
Thus,  the mean-field $T_\mathrm{c}$ is a crossover temperature. 
Figure~\ref{theorySG} (b) shows the mean-field $T_\mathrm{c}^\mathrm{MF}$ and the suppressed $T_\mathrm{c}$~\cite{Yanase}. 
The doping dependence of $T_\mathrm{c}^\mathrm{MF}$ depends on the shape of the Fermi surface. 
 
From the above two theories, one may conclude that 
the difference in the hole doping dependence of $T_\mathrm{s}$ of Hg1201 from Hg1212
can be attributed to the different shape in the basal Fermi surface of Hg1201 and Hg1212. 

The band structure calculations have been performed by the full potential linear muffin-tin orbital method for Hg1201, Hg1212 and Hg1223~\cite{Freeman}. The band calculations indicate the different electronic structures and Fermi surfaces between Hg1201 and the others~\cite{Freeman}. 
The Fermi surface of a single crystal Hg1201 was observed by angle-resolved photoemission measurement~\cite{Lee}.

\section{SPIN FLUCTUATION SPECTRUM}  
\subsection{Scaling}  
\begin{figure}
\begin{center}
\includegraphics[width=0.8\linewidth]{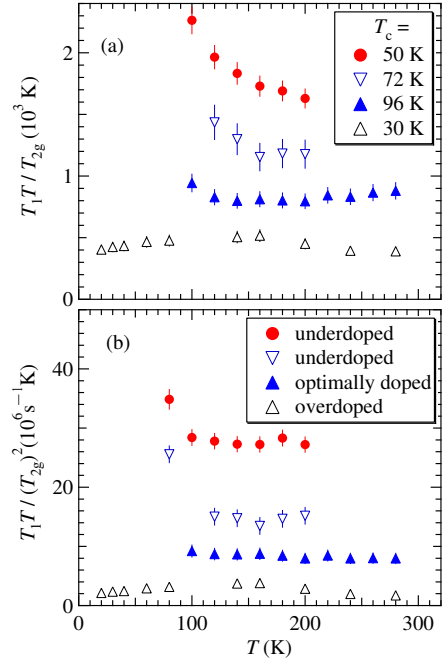}
\end{center}
\caption{
Hole doping dependence of $T_1T/T_{2g}$ (a) and $T_1T/(T_{2g})^2$ (b) for Hg1201.
The data are reproduced from ~\cite{ItohHg1201L,ItohHg1201F,ItohHg1201P}. 
}
\label{Z1Z2Hg1201}
\end{figure}  
\begin{figure}
\begin{center}
\includegraphics[width=0.8\linewidth]{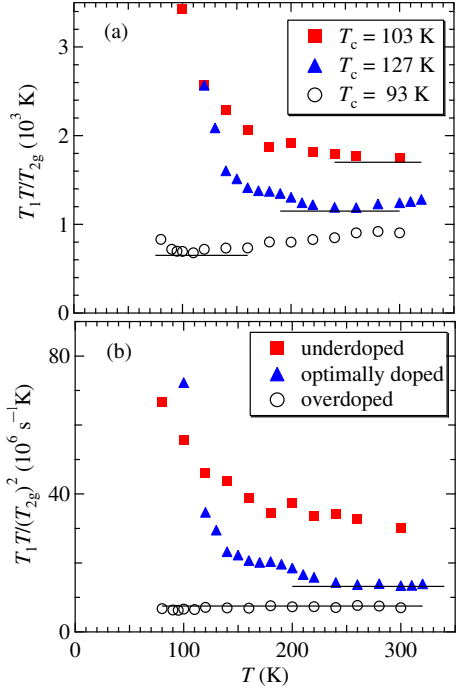}
\end{center}
\caption{
Hole doping dependence of $T_1T/T_{2g}$ (a) and $T_1T/(T_{2g})^2$ (b) for Hg1212.
The data are reproduced from ~\cite{ItohHg1212}. 
Solid lines are eye guides for $T_1T/T_{2g}$ = constant (a) and $T_1T/(T_{2g})^2$ = constant (b). 
}
\label{Z1Z2Hg1212}
\end{figure}  
The ratios of $T_1T$ and $T_{2g}$ tell us the spin fluctuation parameters through Eqs.~(\ref{Z1}),~(\ref{Z2}),~(\ref{SCR1}) and~(\ref{SCR2}). 
In Figs.~\ref{Z1Z2Hg1201} and~\ref{Z1Z2Hg1212}, $T_1T/T_{2g}$ (a) and $T_1T/(T_{2g})^2$ (b) for Hg1201 and Hg1212 are plotted against temperatures from the underdoped to the overdoped samples~\cite{ItohHg1201L,ItohHg1201F,ItohHg1212,ItohHg1201P}. 
The pseudo scaling of $T_1T/T_{2g}$ = constant is observed in the limited temperature region for Hg1212. 
The pseudo scaling temperature region decreases and shifts at lower temperatures by the hole doping. 
At high temperatures, $T_1T/(T_{2g})^2$ = constant holds. 
The value of $T_1T/(T_{2g})^2$ indicates the product of the spin fluctuation amplitude and the spin fluctuation energy. 
Thus, the product of $\chi({\bf Q})\Gamma({\bf Q})$ decreases from the underdoped to the overdoped regimes of Hg1201 and Hg1212. 

Equation~(\ref{Z1}) indicates that $T_1T/T_{2g}$ is proportional to the inverse of the antiferromagnetic correlation length $\xi$ and has a unique scale parameter $T_0$. 
In Figs.~\ref{Z1Z2Hg1201} and~\ref{Z1Z2Hg1212}, $T_1T/T_{2g}$ decreases from the underdoped to the overdoped samples of Hg1201 and Hg1212. 
This is inconsistent with the theoretical doping dependence of the magnetic correlation length near the two dimensional quantum critical point. 
The $\xi^{-1}$ in the SCR theory increases from the weakly antiferromagnetic to the nearly antiferromagnetic regimes and away from the quantum critical point~\cite{MTU}. 
The decrease of $T_1T/T_{2g}$ suggests the decrease of the spin fluctuation energy $T_0$
from the underdoped to the overdoped samples. 
In the overdoped regime, 
the decrease of $T_\mathrm{c}$ can be associated with the decrease of the spin fluctuation energy $T_0$.

In the RPA for the two dimensional $t$-$J$ model,
both $T_1T/T_{2g}$ (a) and $T_1T/(T_{2g})^2$ (b) depend on temperature more or less~\cite{TKF2}.
The value of $T_1T/T_{2g}$ increases with the hole doping~\cite{TKF2},
that is inconsistent with the experimental doping dependence. 
But the decreases of $T_1T/(T_{2g})^2$ with doping is reproduced within the RPA calculations.

In the numerical calculations for the  small size $t$-$J$ model
using the Lanczos diagonalization method, 
the doping dependence of $T_1T/T_{2g}$ below $T$ = 1390 K agrees with the experimental tendency.   
Above $T$ = 2320 K, the doping dependence of $T_1T/T_{2g}$ above $T$ = 2320 K reproduces those of the SCR and the RPA calculations~\cite{JP}. 
 
\subsection{Spin fluctuation parameters}
\begin{figure}
\begin{center}
\includegraphics[width=0.8\linewidth]{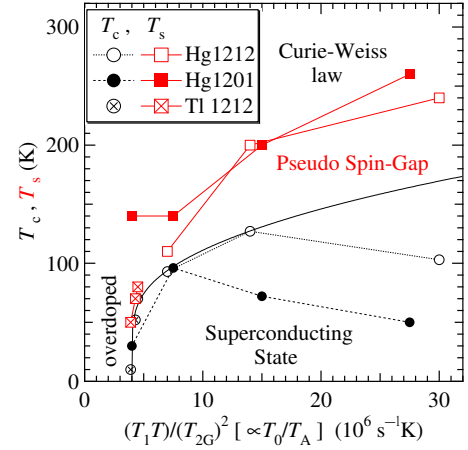}
\end{center}
\caption{
Superconducting transition temperatures $T_\mathrm{c}$ 
are plotted against the spin fluctuation parameters of Hg1201, Hg1212 and Tl1212 (TlSr$_2$CaCu$_2$O$_{7-\delta}$).
Pseudo spin-gap temperatures $T_\mathrm{s}$ 
are also plotted. 
The Cu NMR results of Hg1201 and Hg1212 are reproduced from~\cite{ItohHg1201L,ItohHg1201F,ItohHg1212,ItohHg1201P}.
The Cu NMR results of Tl1212 are reproduced from~\cite{Magishi}. 
}
\label{TcvsSF}
\end{figure}  
Within the framework of Eliashberg-Nambu strong coupling superconductivity theory, 
the actual $T_\mathrm{c}$ 
is determined by competition between the pairing effect and the depairing effect~\cite{OhashiShiba,MillisSC,MonthouxPines}. 
In the spin-fluctuation-mediated superconductors, 
the depairing effect due to low frequency spin fluctuations competes the paring effect due to high frequency ones.
For the $d_{x^2-y^2}$ superconductivity on a square lattice,
the numerical calculation and theoretical consideration tell us that
$T_\mathrm{c}$  is proportional to 
the characteristic energy scale of antiferromagnetic spin fluctuations~\cite{NMU}, 
\begin{equation}
T_\mathrm{c} \propto T_0.    
\label{Tc}
\end{equation}
The spin-fluctuation-induced superconductivity theories are studied for the two and three dimensional Hubbard models~\cite{MU,TMoriya}. 
As a general tendency, 
two dimensional systems have higher $T_\mathrm{c}$ than three dimensional ones~\cite{NMU,Kuroki,Monthoux}. 
Then, the mystery of the high $T_\mathrm{c}$ of the layered compounds is traced back to the two dimensionality and the large scale of the antiferromagnetic spin fluctuation energy.   

Figure~\ref{TcvsSF} shows $T_\mathrm{c}$ plotted against $T_1T/(T_{2g})^2$ of Hg1201, Hg1212,
and Tl2212 reproduced from~\cite{Magishi}. 
If one assumes $T_1T/(T_{2g})^2\propto T_0/T_A$ of Eq.~(\ref{Z2}), 
the linear relation between $T_\mathrm{c}$ and the spin fluctuation energy does not seem to hold. 
The spin fluctuation product $\chi({\bf Q})\Gamma({\bf Q})$ decreases monotonically with the hole doping as in Figs.~\ref{Z1Z2Hg1201} and~\ref{Z1Z2Hg1212}.
This has been recognized in the other systems~\cite{ItohT2g3}. 
Thus, $T_1T/(T_{2g})^2$ is a monotonic function of the doped hole concentration
and then the indicator. 
At some threshold value of $T_1T/(T_{2g})^2$ in the overdoped regime,
$T_\mathrm{c}$ starts to increase toward the optimally doping level. 
Beyond the optimally doping level,  $T_\mathrm{c}$ decreases and $T_\mathrm{s}$ increases  as the hole concentration is reduced. 
The antiferromagnetic spin fluctuation spectrum is different between the underdoped and the overdoped samples. Thus, the overdoped part of the $T_\mathrm{c}$-vs-$T_1T/(T_{2g})^2$ curves indicates 
some correlation between $T_\mathrm{c}$ and the spin fluctuation parameters, 
but the underdoped part indicates the suppression of $T_\mathrm{c}$ due to the grown of the large pseudo spin-gap.     

\subsection{Double-layer coupling: revisited}
\begin{figure}
\begin{center}
\includegraphics[width=1.1\linewidth]{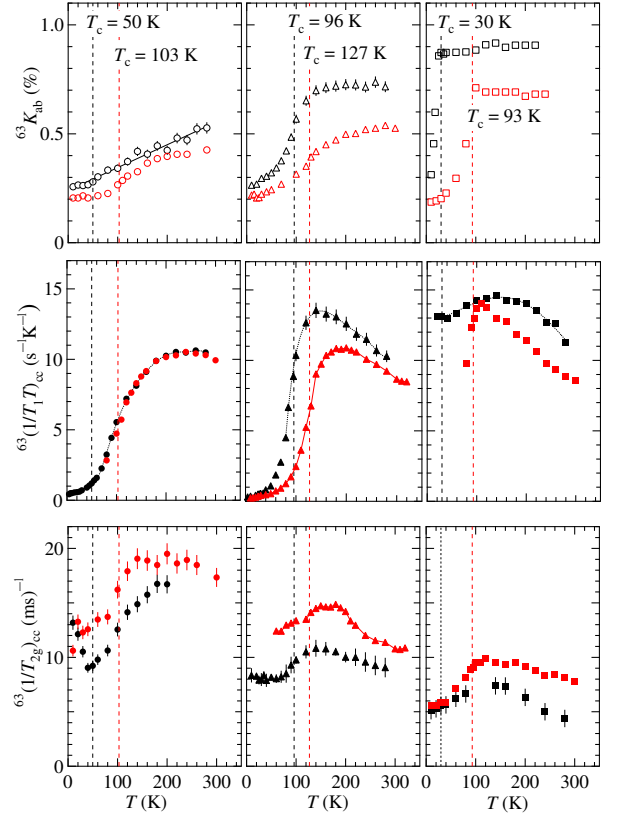}
\end{center}
\caption{
Hg1201 versus Hg1212. 
 $^{63}K_\mathrm{ab}$, $^{63}(1/T_1T)_\mathrm{cc}$, and $^{63}(1/T_\mathrm{2g})_\mathrm{cc}$ 
for Hg1201 (black symbols) and Hg1212 (red symbols) from Figs.~\ref{fig:Hg1201F1} and~\ref{fig:Hg1212F1}~\cite{ItohHg1201L,ItohHg1201F,ItohHg1212,ItohHg1201P}.  
}
\label{Hg1201vsHg1212}
\end{figure}    
From the fact that the uniform spin susceptibility $\chi_s$ is suppressed at low temperatures in
YBa$_2$Cu$_3$O$_{6.6}$ and YBa$_2$Cu$_{4}$O$_{8}$ more than La$_{2-x}$Sr$_{x}$CuO$_4$, the bilayer (double-layer) coupling effect had been proposed to be the primary origin of the pseudo spin-gap~\cite{MM,Ioffe,Millis,Altshuler}. 
The double-layer exchange scattering effect was propsed to account for the different behavior of 1/$T_1T$ and 1/$T_{2g}$ below $T_\mathrm{s}$~\cite{Kishine1,Kishine2}. 
Now the single CuO$_2$ layer Hg1201 is found to possess the pseudo spin-gap in the low-lying excitation spectrum. 
However, the existence of the double-layer coupling was actually confirmed by the neutron scattering~\cite{RM} and NMR~\cite{Matsumura} experiments for low doped YBa$_2$Cu$_3$O$_{6+\delta}$.      
Spin-echo double resonance techniques have been applied to estimate the double-layer coupling constant in Y$_2$Ba$_4$Cu$_7$O$_{15-\delta}$~\cite{Rivo,Suter} and
in Bi$_2$Sr$_2$Ca$_2$Cu$_3$O$_{10}$~\cite{Statt}. 
The angle dependence of the Gaussian decay rate 1/$T_{2g}$ was measured to estimate the like-spin interlayer coupling for Hg1223~\cite{Goto}.   

Figure~\ref{Hg1201vsHg1212} shows $^{63}K_\mathrm{ab}$, $^{63}(1/T_1T)_\mathrm{cc}$, and $^{63}(1/T_\mathrm{2g})_\mathrm{cc}$ for Hg1201 (black symbols) and Hg1212 (red symbols)
from Figs.~\ref{fig:Hg1201F1} and~\ref{fig:Hg1212F1}
to compare two systems~\cite{ItohHg1201L,ItohHg1201F,ItohHg1212,ItohHg1201P}. 
The NMR results of Hg1212 is quantitatively different from those of Hg1201. 

As to the underdoped Hg1201 and Hg1212 samples,
$^{63}(1/T_1T)_\mathrm{cc}$ of Hg1212 is nearly the same as that of Hg1201
but the $T_\mathrm{c}$ is about twice higher than Hg1201.
The different point is the underdoped $^{63}(1/T_\mathrm{2g})_\mathrm{cc}$ of Hg1212 
higher than that of Hg1201. 
Thus, not the low frequency spin fluctutions but the high frequency ones
contribute the higher $T_\mathrm{c}$.  

As to the optimally doped Hg1201 and Hg1212 samples,
the spin part of the Cu Knight shift of Hg1212 and  $^{63}(1/T_1T)_\mathrm{cc}$ are smaller than those of Hg1201,
whereas the $^{63}(1/T_\mathrm{2g})_\mathrm{cc}$ is higher than that of Hg1201. 
The electron spin-spin correlation function $\chi_\mathrm{inter}$ due to the double-layer coupling
is competitive in the in-plane correlation function $\chi_\mathrm{intra}$ of $^{63}K_\mathrm{ab}$ and $^{63}(1/T_1T)_\mathrm{cc}$ but is additive in $^{63}(1/T_\mathrm{2g})_\mathrm{cc}$~\cite{Kishine2}. 
Although the double-layer coupling is not a primary origin of the pseudo spin-gap, 
it surely affects the microscopic magnetic properties.  
 
\section{CONCLUSION}
\label{sec:conclusion}
	The flat CuO$_2$ plane and the large pseudo spin-gap are the characteristics of the mercury-based high-$T_\mathrm{c}$ superconducting cuprates Hg1201 and Hg1212. The role of the pseudo spin-gap in the higher $T_\mathrm{c}$ is still unclear. Is the pseudo spin-gap a consequence of the enhanced superconducting fluctuations to suppress the mean field $T_\mathrm{c}^{\mathrm MF}$ ? Then, we should explore the layered compounds  with higher energy spin fluctuations and any method to suppress the superconducting fluctuations to get higher $T_\mathrm{c}$.      

\acknowledgments
We would like to thank S. Adachi, A. Yamamoto, A. Fukuoka, K. Tanabe, N. Koshizuka, K. Yoshimura, and H. Yasuoka 
for the fruitful collaboration, and Y. Ohashi, J. Kishine, and Y. Yanase for the valuable discussions on theoretical studies.   	 


\end{document}